\documentstyle[11pt]{article}
\begin{document}
\title{Probing Physics of Magnetohydrodynamic Turbulence Using Direct Numerical
Simulation}
\author{M.\ K.\ Verma and G. Dar \\
{\em Department of Physics, Indian Institute of Technology,} \\ {\em Kanpur
208016, INDIA}}
\date{\today }
\maketitle

\begin{abstract}
The energy spectrum and the nolinear cascade rates of MHD turbulence is not
clearly understood. We have addressed this problem using direct numerical
simulation and analytical calculations. Our numerical simulations indicate
that Kolmogorov-like phenomenology with $k^{-5/3}$ energy spectrum, rather
than Kraichnan's $k^{-3/2}$, appears to be applicable in MHD turbulence.
Here, we also construct a self-consistent renomalization group procedure in
which the mean magnetic field gets renormalized, which in turns 
yields $k^{-5/3}$
energy spectrum. The numerical simulations also show that the fluid energy
is transferred to magnetic energy. This result could shed light on the
generation magnetic field as in dynamo mechanism.
\end{abstract}

\newpage\ 

\section{Introduction}

The fluid parcels in a turbulent flow have random motion. However, the
random flow velocities obeys certain properties. Since most of the talks in
the conference dealt with chaos which also yields random signals, it is
important to contrast the difference between chaos and turbulence. Chaos can
occurs in a nonlinear system with a few (3-10) degrees of freedom, but a
turbulent system has many (order millions or more) degrees of freedom. Also,
a turbulent system may be chaotic; may not be chaotic such as in structures
(e.g., vortex street); or it may have both chaos and structure coexisting
with each other.

Turbulence is ubiquitous and is of very practical importance. Some of the
major applications are in mixing, aeroplane and high speed vehicle design,
atmospheric flows and weather prediction, astrophysical objects like stars,
jets, interplanetary medium etc. Even with its wide industrial, practical,
and theoretical importance, the understanding of turbulence is very weak.
There are many empirical laws from experiments and simulations. There are
some phenomenologies, the most famous among these is by Kolmogorov. There
only a few mathematically rigorous calculations, primarily Kraichnan's
direct interaction approximation, calculations based on renormalization
group etc. In this paper we will focus on some of the statistical properties
of velocity and magnetic fields in a turbulent magnetohydrodynamic (MHD)
plasma. We have attempted to review in a concise manner some of the
phenomenological, numerical, analytical work, and observations from the
solar wind, with an emphasis on our work (with D. A. Roberts, M. L.
Goldstein, J. K. Bhattacharjee, V. Eswaran).

Outline of the paper is as follows. Since the existing MHD turbulence
phenomenologies are motivated by Kolmogorov's fluid turbulence
phenomenology, we review Kolmogorov's arguments in section 2. In section 3
we review the existing MHD turbulence phenomenologies. Section 4 contains
the numerical results, which are compared with the predictions of the
turbulence phenomenologies. In section 5 we briefly report the cascade rates
of fluid and magnetic energies. Section 6 contains a renormalization group
scheme which provides a self-consistent procedure to obtain the effective
mean magnetic field. Section 7 contains conclusions.

\section{Fluid Turbulence}

The fluid flows are described by Navier-Stokes equation, which is 
\begin{equation}
\label{NS}\left( \frac \partial {\partial t}+{\bf u}({\bf x,}t)\cdot
\bigtriangledown \right) {\bf u}({\bf x,}t)=-\frac 1\rho \bigtriangledown p(%
{\bf x},t)+\nu \nabla ^2{\bf u}({\bf x,}t)
\end{equation}
where ${\bf u}({\bf x,}t)$ is the velocity field, $p({\bf x},t)$ is the
pressure field, $\rho $ is the density, and $\nu $ is the viscosity of the
fluid. We assume that the fluid is incompressible, which implies that $%
\bigtriangledown \cdot {\bf u}({\bf x,}t)=0$. The above equation can be made
non-dimensional by scaling the variables by large length, velocity, and time
scales ($L,U,T$). The ratio of the nonlinear terms (II and III term) and the
viscous term (IV term) turns out to be $UL/\nu ,$ which is called the
Reynolds number. When the Reynolds number $Re$ is large (greater that 10,000
or so), we say that the flow is turbulent. Please note that unlike chaos,
there is no critical Reynolds number above which the flow becomes turbulent;
experiments exhibit transitions at different $Re$'s.

Analytic solution to the above equation is not known in turbulent regime.
The existence and uniqueness of the solution itself is not clear at this
moment. From the experiments it is known that the velocity field ${\bf u(x,}%
t)$ is random. There is a range of scales present in the flow between the
energy feeding scale $(L)$ and the dissipation scale $(\eta )$. This
intermediate range is called the inertial range. Kolmogorov conjectured that
the physics in this range does not depend on the energy feeding scale or the
dissipation scale; the energy distribution is homogeneous and isotropic; the
interaction between the velocity modes is local in the Fourier space; and
the energy cascade rate $\Pi (k)$ is independent of $k$ under steady state.
Using these assumptions, the energy spectrum can be deduced immediately
using dimensional analysis. We find that one dimensional energy spectrum $%
E(k)$ ($\int E(k)dk=$ total energy) is given by 
\begin{equation}
\label{Kolmfl}E(k)=K_{Ko}\Pi ^{2/3}k^{-5/3}
\end{equation}
where $K_{Ko}$ is an universal constant called Kolmogorov's constant, $k$ is
the wavenumber. Note that the nonlinear energy cascade $\Pi (k)$ is equal to
the dissipation rate and the energy supply rate of the fluid.

Experiments and simulations \cite{exptsim} are in reasonable agreement with
the above phenomenology. There is a small deviation of the exponent from 5/3
which is attributed to intermittency phenomenon. The Kolmogorov
phenomenology described above is well supported by the calculations based on
Direct interaction approximations \cite{Leslie}, renormalization group
techniques \cite{RG,Mac}, self-consistent mode coupling etc.

Regarding dependence on dimensionality, a brief remark is in order here. In
two dimensions, there are two different powerlaw exponents, 5/3 for small
wavenumber and 3 for large wavenumber regimes. The origin of the exponent 3
is because of enstrophy conservation. We will not discuss this issue here.

After this brief introduction to fluid turbulence, now we turn to MHD
turbulence.

\section{MHD Turbulence}

The incompressible MHD equations are 
\begin{equation}
\label{MHDz}
\begin{array}{c}
\left( \frac \partial {\partial t}\mp 
{\bf B}_0\cdot \bigtriangledown +{\bf z}^{\mp }\cdot \bigtriangledown
\right) {\bf z}^{\pm }=-\frac 1\rho \bigtriangledown p_{tot}+\nu _{+}\nabla
^2{\bf z}^{\pm }+\nu _{-}\bigtriangledown ^2{\bf z}^{\mp } \\ 
\bigtriangledown \cdot {\bf z}^{\pm }=0 
\end{array}
\end{equation}
where ${\bf z^{\pm }=u\pm b.}$ Here {\bf u} is the velocity fluctuation, 
{\bf b} is the magnetic field fluctuation in velocity units (scaled by $%
(4\pi \rho )^{-1/2}$) , ${\bf B_0}$ is the mean magnetic field in velocity
units, and $\nu _{\pm }=(\nu \pm \lambda )/2,$where $\lambda $ is the
resistivity. The field $p_{tot}$ is sum of thermal and magnetic pressure.

The Alfv\'en waves are the basic modes of incompressible MHD equations. In
absence of the nonlinear term $({\bf z}^{\mp }\cdot \bigtriangledown ){\bf z}%
^{\pm }$, ${\bf z}^{\pm }$ are the two independent modes travelling
antiparallel and parallel to the mean magnetic field. However, when the
nonlinear term is present, new modes are generated. These modes interact
with each other, which results in a turbulent behaviour of the fluctuations.

Magnetohydrodynamic (MHD) turbulence is more complex than the fluid
turbulence. There are several MHD turbulence phenomenologies. In the
following discussion we will state the arguments of Dobrowolny et al. \cite
{mhdkraich} from which we can obtain the existing MHD\ turbulence
phenomenologies. The nonlinear interactions modify the amplitudes of the
eddies. We assume that the interaction between the fluctuations are local in
wavenumber space, and that in one interaction, the eddies $z_k^{\pm }$
interact with the other eddies of similar sizes for time interval $\tau
_k^{\pm }$. Then from Eq. (\ref{MHDz}), the variation $\delta z_k^{\pm }$ in
the amplitudes of these eddies during these interval is given by 
\begin{equation}
\delta z_k^{\pm }\approx \tau _k^{\pm }z_k^{+}z_k^{-}k. 
\end{equation}
In $N$ such interactions, because of their stochastic nature, the amplitude
variation will be $\Delta z_k^{\pm }\approx \sqrt{N}(\delta z_k^{\pm })$.
Therefore, the number of interactions $N^{\pm }$ it takes to obtain a
variation equal to its initial amplitude $z_k^{\pm }$ is 
\begin{equation}
N^{\pm }\approx \frac 1{k^2\left( z_k^{\mp }\right) ^2\left( \tau _k^{\pm
}\right) ^2} 
\end{equation}
and the corresponding time $T^{\pm }=N\tau _k^{\pm }$ is 
\begin{equation}
T^{\pm }\approx \frac 1{k^2\left( z_k^{\mp }\right) ^2\tau _k^{\pm }}. 
\end{equation}
The time scale of the energy transfer at wavenumber $k$ is assumed to be $%
T^{\pm }$. Therefore, the fluxes $\Pi ^{\pm }$ of the fluctuations $z_k^{\pm
}$ can be estimated as 
\begin{equation}
\Pi ^{\pm }\approx \frac{\left( z_k^{\mp }\right) ^2}{T^{\pm }}\approx \tau
_k^{\pm }\left( z_k^{\pm }\right) ^2\left( z_k^{\mp }\right) ^2k^2. 
\end{equation}

By choosing different interaction time-scales, one can obtain different
energy spectra. Kraichnan (1965) and Dobrowolny et al. (1980) \cite
{mhdkraich} argued that the interacting $z_k^{+}$ and $z_k^{-}$ modes will
get separated because of the mean magnetic field in one Alfv\'en time-scale.
Therefore, they chose Alfv\'en time scale $\tau _A=(kB_0)^{-1}$ as the
relevant time-scale and found that 
\begin{equation}
\label{Dobro}\Pi ^{+}\approx \Pi ^{-}\approx \frac
1{B_0}E^{+}(k)E^{-}(k)k^3=\Pi .
\end{equation}
If we assume that $E^{+}(k)\approx E^{-}(k),$ then we immediately obtain 
\begin{equation}
E^{+}(k)\approx E^{-}(k)\approx \left( B_0\Pi \right) ^{1/2}k^{-3/2}
\end{equation}
In absence of mean magnetic field, the magnetic field of the largest eddy
was taken as $B_0$. Kraichnan (1965) also argued that the fluid and magnetic
energies are equipartitioned. We refer the above phenomenology either as one
due to Dobrowolny et al. or the generalized Kraichnan's phenomenology.

If the nonlinear time-scale $\tau _{NL}^{\pm }\approx (kz_k^{\mp })$ is
chosen as the interaction time-scales for the eddies $z_k^{\pm }$, we obtain 
\begin{equation}
\Pi ^{\pm }\approx \left( z_k^{\pm }\right) ^2\left( z_k^{\mp }\right) k,
\end{equation}
which in turn leads to 
\begin{equation}
\label{eq:mhd}E^{\pm }(k)=K^{\pm }(\Pi ^{\pm })^{4/3}(\Pi ^{\mp
})^{-2/3}k^{-5/3},
\end{equation}
where $K^{\pm }$ are constants, which we will refer to as Kolmogorov's
constants for MHD turbulence. Because of its similarity with Kolmogorov's
fluid turbulence phenomenology, we refer to this phenomenology as
Kolmogorov-like MHD turbulence phenomenology. This phenomenology was first
given by Marsch \cite{Marsch}. It is also a limiting case of a more
generalized phenomenology constructed by Matthaeus and Zhou, Zhou and
Matthaeus \cite{ZM}, which is 
\begin{equation}
\Pi ^{\pm }=\frac{A^2E^{+}(k)E^{-}(k)k^3}{B_0+\sqrt{kE^{\pm }(k)}}
\end{equation}
where $A$ is a constant. Here the small wavenumbers ($\sqrt{kE^{\pm }(k)}\gg
B_0$) follow 5/3 spectrum, whereas the large wavenumbers ($\sqrt{kE^{\pm }(k)}%
\ll B_0$) follow 3/2 spectrum.

After the discussion on these phenomenologies one would naturally ask which
of the phenomenologies is applicable in the solar wind. From the arguments
by Matthaeus and Zhou, and Zhou and Matthaeus \cite{ZM} we see that
Kraichnan or Dobrowolny et al.'s phenomenology is expected to hold when $%
B_0\gg \sqrt{kE^{\pm }(k)}$; on the other hand Kolmogorov-like phenomenology
is expected to be applicable when $B_0\ll \sqrt{kE^{\pm }(k)}$. We would
like to test these scaling arguments from the solar wind observations and
simulations. In the solar wind, which is a good test ground for MHD
turbulence theories, we find that the exponent of the total energy is $%
1.69\pm 0.08$, whereas the exponent of the magnetic energy is $1.73\pm 0.08$ 
\cite{MG82}, somewhat closer to 5/3 than 3/2. This is more surprising
because $B_0\gg \sqrt{kE^{\pm }(k)}$ for inertial range wavenumbers in the
solar wind. Also, in the solar wind we do not find a break from 5/3 to 3/2
spectrum as predicted by Matthaeus and Zhou, Zhou and Matthaeus \cite
{ZM,Verma:sim}. Hence, from the comparison with the solar wind observations,
it appears that there is some inconsistency in the phenomenological
arguments given above. To explore these issues further, we performed
numerical simulations and some analytic studies. As we will described below,
the numerical simulations also tend to indicate that the Kolmogorov-like
phenomenology, rather than Kraichnan's or Dobrowolny et al.'s phenomenology,
is probably applicable in MHD turbulence.

We have applied renormalization groups to analyse the above equations. We
find that under certain assumptions, $B_0$ gets renormalized as we go from
large length-scales to smaller length-scales. In other words, $B_0$
appearing in the Kraichnan's or Dobrowolny et al.'s argument must be $k$
dependent. This leads to $k^{-5/3}\,$energy spectra, which appears to be
consistent with the solar wind observations and the simulation results. We
will describe these ideas in more detail in the following sections.

Before we proceed further, we point out that the normalized cross helicity $%
\sigma _c$, defined as $(E^{+}-E^{-})/(E^{+}+E^{-})$, and the Alfv\'en ratio 
$r_A$, defined as the ratio of fluid energy and magnetic energy, play an
important role in MHD turbulence.

\section{Numerical Simulation}

To probe numerically the physics at the inertial range, one usually solves
the MHD equations using the pseudospectral method with introduction of
hyperviscosity and hyperrestivity. For large resolution turbulence
simulations, spectral method is preferred over finite difference methods
because the derivatives can be calculated accurately using the Fourier
transforms. Note that any error in these equations could propagate very
fast, which could make the simulation unreliable. Most of our runs were
performed in two-dimensions because of the expense associated with large
three-dimensional runs. Fortunately, the scaling arguments described in the
earlier section does not depend on the dimensionality because the absolute
equilibrium theories predict a forward cascade of total energy in both two
and three dimensions. The simulation method is as follows.

The equations are time advance in Fourier space. The nonlinear terms are
calculated in real space, then its Fourier transform is calculated using
FFT. For details of the simulation, refer to Biskamp and Welter \cite{MHDsim}
and Verma et al. \cite{Verma:sim}. Earlier high resolution simulations were
performed by Biskamp and Welter (1989), Pouquet et al. (1988), and Politano
et al. (1989) \cite{MHDsim}. In all these simulations there was no strong
evidence for either 3/2 or 5/3 spectral evidence.

We have taken $\nu =\lambda ,$ (or $\nu _{-}=0$) in our simulation. The
viscous term has been modified by an addition of hyperviscous and
hyperresistive term. The modified viscous term is 
\begin{equation}
\nu \left( \bigtriangledown ^2+\frac{\bigtriangledown ^4}{k_{eq}^2}\right) 
{\bf z}^{\pm }. 
\end{equation}
Our $\nu =5.0\times 10^{-6}$, but $k_{eq}=10.0$ which yields Reynolds number
of approximately $2\times 10^5$ at large scales and 500 at small scales. The
maximum resolution of our simulation in 2D is $512^2$. The parameter $%
dt=5.0\times 10^{-4}.$ The runs we consider have $\left( B_0,\sigma
_c\right) =(0,0),(0,0.25),(0,0.9),(1,0),(1,0.25),(1,0.9),(5,0.25),(5,0.9).$
The $\sigma _c$ values are for $T=0$. We have taken the initial Alfv\'en
ratio to be 1.0. The simulation were initialized with a box spectrum out to $%
k=4$ for $B_0=0$, and a $k^{-1}$ spectrum out to $k=15$ for $B_0=5.0$.

We find that after approximately $T=10$ the system has reached a fully
developed turbulent state, i.e., the energy spectrum shows approximate power
laws in the inertial range. We obtained the spectral indices by fitting a
straight line in the inertial range, and found that it was difficult to
distinguish between the indices 3/2 and 5/3. Therefore, we decided to study
the energy cascade rates $\Pi ^{\pm }\,$ leaving a sphere of radius $K$ in
the wavenumber space which is given by 
\begin{equation}
\Pi ^{\pm }(K)=-\frac \partial {\partial t}\int_0^KE^{\pm }(k,t)dk-2\nu
\int_0^K[1+(k/k_{eq})]^2k^2E^{\pm }(k)dk.
\end{equation}
The cascade rates $\Pi ^{\pm }(K)$ were calculated numerically and compared
with the predictions of the phenomenologies. The Kolmogorov-like
phenomenology predicts that $(E_k^{-}/E_k^{+})/(\Pi ^{-}/\Pi
^{+})^2=K^{-}/K^{+}$, while the Dobrowolny et al.'s phenomenology predicts
that $\Pi ^{+}=\Pi ^{-}$. In our simulation we find that for initial $\sigma
_c=0$ and 0.25, $(E_k^{-}/E_k^{+})/(\Pi ^{-}/\Pi ^{+})^2\approx 1$. For
large $\sigma _c$ this equality does not hold. Note that uncertainties is
quite large in our simulations. However, we found in all our runs that the
larger of $E_k^{+}$ and $E_k^{-}$ had larger cascade rate, thus negating
Dobrowolny et al.'s predictions. In our simulations the mean magnetic field
(or the magnetic field of the largest eddies) were at least $\approx 1$,
whereas the amplitudes of the largest $z^{\pm }$ fluctuations in the
inertial range $(k>10)$ were 0.2 to 0.5 (initial total energy = 1). Hence $%
z^{\pm }<B_0$, and generalized Kraichnan's model should have been
applicable, but we find otherwise. Hence we conclude that our numerical
result are contrary to the predictions of the phenomenologies. The
Kolmogorov-like phenomenology appears to hold for small $\sigma _c$ (with $%
K^{+}=K^{-}$) in all $B_0=0-5$ regime. However, for large $\sigma _c$, there
appears to be inconsistency. It is possible that the Kolmogorov-like
phenomenology may still hold for large $\sigma _c$ with constants $K^{+}$
and $K^{-}$ being unequal and dependent on $\sigma _c$. This issue is under
current investigation. (The details of the above simulation results are
described in Verma et al. \cite{Verma:sim}).

\section{Cascade rates of magnetic and fluid energies}

The origin of magnetic field in the earth, sun, and in the universe is an
important problem. It is generally believed that the fluid energy gets
transferred to the magnetic energy due to nonlinear interactions, a
mechanism know as dynamo. There are many models to get the desired magnetic
field configuration in the astrophysical objects. Here we have attempted to
investigate the cascade rates of magnetic and fluid energies. These studies
will shed light on the energy transfer mechanisms involved in dynamo
mechanism. Since this work is in progress, here we are reporting only the
preliminary results.

The equations for fluid and magnetic energies are as follows: 
\begin{equation}
\label{u2}\left( \frac \partial {\partial t}-2\nu k^2\right) u^2=-{\bf %
u\cdot \bigtriangledown }p-{\bf u\cdot }\left[ {\bf u\cdot \bigtriangledown u%
}\right] +{\bf u\cdot }\left[ {\bf b\cdot \bigtriangledown b}\right] 
\end{equation}
\begin{equation}
\label{b2}\left( \frac \partial {\partial t}-2\lambda k^2\right) b^2={\bf %
b\cdot }\left[ {\bf b\cdot \bigtriangledown u}\right] -{\bf b\cdot }\left[ 
{\bf u\cdot \bigtriangledown b}\right] 
\end{equation}
The cascade rates by evaluating $-\int_0^KT(k)dk$ where $T(k)$'s are the
nonlinear terms appearing in the right hand side of the above equations. The
pressure term does not yield any energy transfer. The second term in the RHS
of Eq. (\ref{u2}) yields the energy transfer from $u\rightarrow u$ $(\Pi
_{u})$, while the third term yields energy transfer $%
u\rightarrow b(\Pi_{u\rightarrow b})$. The terms in the RHS\ of Eq. (\ref
{b2}) primarily yield the energy transfer from $b\rightarrow u$ $(\Pi
^{total}_{b})$(the transfer from $b\rightarrow b$ has been ignored). The 
total energy transfer rates coming from {\bf u} and {\bf b} spheres
of radius $K$ are $\Pi_{u}^{total}$ and $\Pi_{b}^{total}$ respectively.
We have obtained these cascade rates numerically.

We solve the MHD equations numerically one a $256\times 256$ grid using the
method described in the earlier section. We choose $\nu =\lambda =10^{-5}$, $%
k_{eq}=10$, and initial $\sigma _c=0,r_A=1.0$ . Note that there is no
forcing in our simulation. We have carried out our simulation till $T=7.5$.
On DEC 2000 4/233 the computer time required to reach $T=7.5$ was
approximately 4 hours. At $T=7.5$ we have calculated the cascade rates by
calculating the triple correlations. These cascade rates are shown in Figure
1. We find that $\Pi_{u\rightarrow b}$ is positive, while $\Pi
_{b}$ negative. At $k_{\max }$ we have $\Pi_{u\rightarrow
b}=- \Pi_{b}^{total}$ Hence, there is a net transfer of energy from
the fluid energy to the magnetic energy. The energy evolution studies
indicate that the total fluid energy $(u^2/2)$ is decaying, while the
magnetic energy $(b^2/2)$ is almost constant in time. Hence, the dissipation
rate of magnetic energy appears to be balanced by the energy gained from the
transfer rate from the fluid energy. It is to be seen when the magnetic
energy get enhanced. In any case, these simulations show that there is a net
transfer of energy from velocity field to magnetic field. We also notice
that the energy cascade rate $\Pi_{u}$ is much smaller that
both $\Pi_{u\rightarrow b}$ and $\Pi_{b}^{total}$. Careful analysis
of the cascade rates for various parameter range is in progress.

We have attempted to obtain the perturbative solutions of MHD equations.
This is described below.

\section{Kolmogorov-like powerlaw using renormalization groups}

There have been earlier attempts of applying renormalization groups \cite
{MHDrg} and other analytic technique, e.g.
Eddy-Damped-Quasi-Normal-Markovian approximation, to MHD\ turbulence. In
earlier renormalization group calculations, corrections to viscosity and
diffusivity are evaluated by the nonlinear terms in presence of forcing. In
our calculation presented here, we attempt to calculate corrections to the
mean magnetic field due to the nonlinear terms.

The basic idea of our calculation is to obtain the effective $B_0$ at higher
wavenumbers. In the phenomenology of Kraichnan and Dobrowolny et al. \cite
{mhdkraich} the external field or the magnetic field of the largest eddy is
taken to $B_0$. Here we construct a self-consistent scheme in which the
effective $B_0$ is taken to be the magnetic field of the next-largest eddy.
For example, for Alfv\'en waves of wavenumber $k$, the effective magnetic
field $B_0$ will be the magnetic field of the eddy of size $k/10$ or so.
This argument is based on the physical intuition that for the scattering of
the Alfv\'en waves at a wavenumber $k$, the effects of the magnetic field of
the next-largest eddy is much more than that of the external field. (For
similarity, please note that in WKB method , local inhomogeneity of the
medium determines the amplitude and phase evolution). We will find in the
following discussion that the $k$ dependent $B_0$ yields $k^{-5/3}$ energy
spectra. For simplicity we have taken $E^{+}(k)=E^{-}(k)$ and $r_A=1$.

The MHD equation in the Fourier space is \cite{mhdkraich} 
\begin{equation}
\label{mhdeqn}\frac d{dt}z_i^{\pm }({\bf k},t)\mp i\left( {\bf B}_0\cdot 
{\bf k}\right) z_i^{\pm }({\bf k},t)=
-i M_{ijm}({\bf k})\int d{\bf p}z_j^{\mp }({\bf p},t)z_m^{\pm }({\bf k-p},t) 
\end{equation}
where 
\begin{equation}
M_{ijm}({\bf k})=k_jP_{im}({\bf k});\ P_{im}({\bf k})=\delta _{im}-\frac{%
k_ik_m}{k^2}, 
\end{equation}
Here we have ignored the viscous terms. The above equation will, in
principle, yield an anisotropic energy spectra (different spectra along and
perpendicular to ${\bf B}_0$). Here we modify the above equation in the
following way to preserve isotropy 
\begin{equation}
\label{mhdk}
\frac d{dt}z_i^{\pm }({\bf k},t)\mp i\left( B_0k\right) z_i^{\pm
}({\bf k},t)=-i M_{ijm}({\bf k})\int d{\bf p}z_j^{\mp }({\bf p},t)z_m^{\pm }
({\bf k-p},t) 
\end{equation}
This equation can be thought of as an effective equation with an isotropic
random mean field.

The calculation of $B_0(k)$ using the renormalization group is done by a
procedure adopted by McComb, McComb and Watt, Zhou et al., and others \cite
{Mac}. The wavenumber range $(k_0..K_N)$, is divided logarithmically into $N$
divisions. The $n$th shell is $(k_{n-1}..k_n)$ where $k_n=s^nk_o(s>1)$.
Brief summary of the RG operation used is given below (For details of the
procedure, refer to \cite{Mac} and \cite{Verma:rg}).

\begin{enumerate}
\item  Decompose the modes into the modes to be eliminated $(k^{<})$ and the
modes to be retained $(k^{>}).$ In the first iteration $(k_0..k_1)=k^{<}$
and $(k_1..k_N)=k^{>}$.

\item  We rewrite the Eq. (\ref{mhdk}) for $k^{<}$ and $k^{>}$. The equation
for the retained modes are 
\begin{equation}
\label{MHDk}
\begin{array}{c}
\frac{d}{dt}z_i^{\pm >}({\bf k},t)\mp i\left( B_0(k)k\right) 
z_i^{\pm >}({\bf k},t)=-i M_{ijm} ({\bf k}) \int d{\bf p}\left[ 
z_j^{\mp >}({\bf p},t)z_m^{\pm >}({\bf k-p},t)\right] + \\ 
\left[ z_j^{\mp >}({\bf p},t)z_m^{\pm <}({\bf k-p},t)
+z_j^{\mp <}({\bf p},t)z_m^{\pm >}({\bf k-p},t)+z_j^{\mp <}
({\bf p},t)z_m^{\pm <}({\bf k-p},t)\right] 
\end{array}
\end{equation}
We get a similar equation for $z_i^{\pm <}({\bf k},t)$ modes

\item  The terms given in the second bracket in the RHS of Eq. (\ref{MHDk})
is calculated perturbatively and the $z_i^{\pm <}({\bf k},t)$ modes are
averaged out. The effective equation after this operation is 
\begin{equation}
\begin{array}{c}
\frac{d}{dt}z_i^{\pm >}({\bf k},t)\mp i
\left( \left[ B_0(k)+\delta B_0^{\pm }(k)\right] k\right) 
z_i^{\pm >}({\bf k},t) \\ 
=-i M_{ijm}({\bf k})\int d{\bf p}\left[
z_j^{\mp >}({\bf p},t)z_m^{\pm >}({\bf k-p},t)\right] 
\end{array}
\end{equation}
where 
\begin{equation}
\delta B_0^{\pm }(k)=-\frac 12\int_{{\bf p+q=k}}d{\bf q}b_2(k,p,q)\left( 
\frac{E(q)}{4\pi q^2}\right) \left( \frac 1{pB_0(p)+qB_0(q)}\right) 
\end{equation}
The term $b_2(k,p,q)=kp(1+z^2)(z+xy),$where $x,y,z$ are the cosine of angles
between $({\bf p,q),(q,k),}$ and $({\bf k,p)}$ \cite{Leslie}. Note that $%
\delta B_0^{+}(k)=\delta B_0^{-}(k)$ (for $E^{+}=E^{-}$ and $r_A=1$). Let us
denote 
\begin{equation}
B_1(k)=B_0(k)+\delta B_0(k)
\end{equation}

\item  We keep eliminating the shells by the above procedure. After $n+1$
iterations we obtain 
\begin{equation}
\label{rg1}
B_{n+1}(k)=B_n(k)+\delta B_n(k)
\end{equation}
where 
\begin{equation}
\label{rg2}\delta B_n(k)=-\frac 12\int_{{\bf p+q=k}}d{\bf q}b_2(k,p,q)\left( 
\frac{E(q)}{4\pi q^2}\right) \left( \frac 1{pB_n(p)+qB_n(q)}\right) 
\end{equation}

\item  We substitute the following forms for $E(k)$ and $B_n(k)$ in the Eqs.
(\ref{rg1},\ref{rg2}) and solve for $B_n^{*}(k^{\prime })$ numerically \cite
{Mac,Verma:rg}. 
$$
\begin{array}{c}
E(k)=\alpha \Pi ^{2/3}k^{-5/3} \\ 
B_n(k_nk^{\prime })=\alpha ^{1/2}\Pi ^{1/3}k_n^{-1/3}B_n^{*}(k^{\prime })
\end{array}
$$
We start with initial $B_0(k_i)$ at each shell, and keep iterating till $%
B_{n+1}^{*}(k^{\prime })\approx B_n^{*}(k^{\prime })$, that is, till the
solution converges. We find that $B_n^{*}(k^{\prime })\ $is approximately
proportional to $k^{\prime -1/3}$. Hence, the mean magnetic field scales as $%
k^{-1/3}$, and the energy spectra scales as $k^{-5/3}$ in this
self-consistent scheme.
\end{enumerate}

Hence, we see that scaling of $B_0$ leads to $k^{-5/3}$ energy spectra. This
is a self-consistent scheme that shows the plausibility of Kolmogorov-like
energy spectra for MHD\ turbulence. Note that simulations and the solar wind
observations appear to favour the Kolmogorov-like phenomenology. Therefore,
the above analytical analysis is a promising step toward a proper
understanding of the statistical theory of MHD\ turbulence.

\section{Conclusions}

In this paper we have reviewed some of the phenomenological, observational,
numerical, and analytic work in the statistical theory of MHD turbulence,
with emphasis on our studies. We find that the solar wind observations and
the numerical results are inconsistent with the predictions of the existing
MHD turbulence phenomenologies. The energy spectrum and the cascade rates
appear to be closer the predictions of Kolmogorov-like phenomenology even
when the mean magnetic field or the magnetic field of the largest eddies is
large compared to the inertial range velocity and magnetic field
fluctuations (a region where Kolmogorov-like theory is not expected to hold).

We have attempted to obtain Kolmogorov-like energy spectrum in MHD
turbulence in presence of arbitrary $B_0$ by postulating that the effective $%
B_0$ is scale dependent, unlike what has been taken by Kraichnan and
Dobrowolny et al. We have constructed a renormalization group scheme and
shown that the self consistent $B_0(k)\propto k^{-1/3}$ and $E(k)\propto
k^{-5/3}$. This analysis has been worked out when $E^{+}=E^{-}$ and $r_A=1.$
The generalization to arbitrary parameters, or at least to the limiting
cases, is also planned. We will be able to the get the Kolmogorov's
constants for MHD\ turbulence analytically using this procedure; these
constants will be useful for the large-eddy-simulations (LES) of MHD
turbulence.

We have also carried out a preliminary study of the cascade rates of fluid
and magnetic energies. We find that there is a net transfer of fluid energy
to magnetic energy. We are investigating the parameter regimes in which
these cascade rates are large enough (more than dissipation rates) so that
the total magnetic energy increases with time, what is found in dynamo
theory.

Lastly, we would like to mention that the MHD turbulence theories are very
important for modelling various astrophysical phenomena and plasma
processes. We have estimated turbulent heating, nonclassical viscosity and
resistivity of the solar wind using the MHD turbulence phenomenologies \cite
{Verma:sw}. In this light, search for a satisfactory theory of MHD
turbulence appears quite important.

We thank all our collaborators, D. A. Roberts, M. L. Goldstein, J. K.
Bhattacharjee, and V. Eswaran. MKV\ also thanks V. Subrahmanyam, M. Barma,
V. Ravishankar, and D. Sa for numerous useful discussions.

\vspace{2.0cm}
\begin{center}
{\bf FIGURE CAPTIONS}
\end{center}
\vspace{0.5cm}
\noindent Figure 1  This figure shows the fluxes for a simulation with initial
$\sigma_c=0.$ and $r_A=1.0$. The fluxes are found at $t=7.5$ when 
$\sigma_c=0.015$ and $r_A=0.4$. 

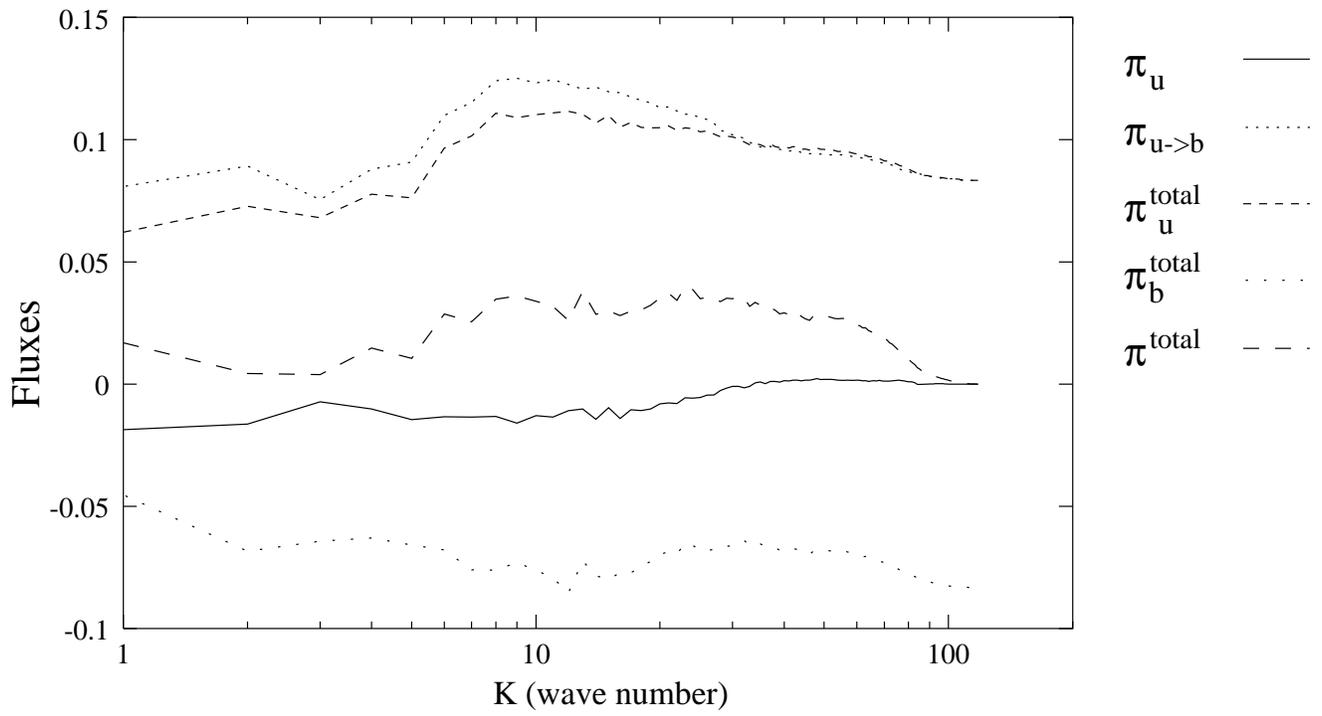
\end{document}